# ACHIEVING SENSOR FUSION FOR COLLABORATIVE MULTI-LEVEL MONITORING OF PIPELINE INFRASTRUCTURES

Grigore STAMATESCU[1]

*Sistemele de monitorizare pe scară largă permit colectarea eficientă a datelor din câmp la rezoluții temporale și spațiale ridicate. Un exemplu relevant îl constituie aplicarea acestor sisteme în domeniul monitorizării infrastructurilor de transport al produselor petroliere pentru prevenirea scurgerilor și a accesului neautorizat, ce pot cauza daune de mediu și economice semnificative. Lucrarea discută o arhitectură de sistem pentru colectarea și prelucrarea datelor bazată pe integrarea colaborativă a rețelelor de senzori wireless și a platformelor aeriene autonome. Trei metode de fuziune senzorială: filtrajul Kalman, Fuzzy Sensor Validation and Fusion (FUSVAF) și prelucrarea bazată pe consens, sunt evaluate în vederea reducerii inteligente a cantității de date și conștientizarea de nivel înalt a situației, cu eliminarea blocajelor de comunicație în rețeaua multi-nivel. Sunt prezentate rezultate obținute prin simulare și experimental.*

*Large scale monitoring systems enable efficient field level data collection at high temporal and spatial resolutions. One example is the deployment of such systems in pipeline infrastructure applications which have to be monitored for leaks and protected from unauthorized access, with the potential of causing significant environmental and economic damage. The paper discusses a multi-level system architecture for data collection and processing based on the collaborative integration of wireless sensor networks and unmanned aerial vehicles. Three sensor fusion methods: Kalman filtering, Fuzzy Sensor Validation and Fusion (FUSVAF) and Consensus-based processing, are considered for intelligent data reduction and situational awareness while alleviating communication bottlenecks across the multi-level network. Simulation and experimental results are presented.*

**Keywords**: wireless sensor networks, unmanned aerial vehicles, collaborative monitoring, sensor integration, sensor fusion.

## 1. Introduction

The fundamental role played by embedded networked systems in the Industrial- IoT, prominently represented by wireless sensor networks, with well established constraints on local computing and communication, is leading to the emergence and wide spread adoption of new ubiquitous monitoring and control technologies. The applications of these new technologies to solve social, economic and environmental challenges, are to be found in areas like security: critical infrastructure systems protection, energy: future smart grids, advanced metering infrastructures, industry: industrial wireless sensor networks (IWSN) or healthcare: body sensor networks and ambient assisted living and care for the elderly and chronic disease patients. Such a dense space-temporal instrumentation of the physical world leads to huge quantities of raw data and challenges that need to be addressed for effective and secure management and operation of such systems. A first step is to define efficient architectures and topologies for distributed information processing which assure the transformation of collected data to higher level pieces of information. Recent developments in embeded networked sensing and unmanned aerial vehicles, along with standardization efforts have increased the adoption by industry users in real-world smart, safe and sustainable infrastructure systems.

---

[1]Asst. Prof., Dept. of Automatic Control and Industrial Informatics, University „Politehnica" of Bucharest, e-mail: grigore.stamatescu@upb.ro

One approach to solve these ongoing challenges in an industrial system is the application of multi-sensor data fusion (MSDF) as a complex solution for dealing with incomplete, uncertain information coming from heterogeneous data sources in security-constrained environments. A defining problem for the domain is that of detecting, localization, tracking and classification of events, enabling higher level decision entities to act in order to optimize the overall system operation or to minimize losses in the case of failure. Data integration is carried out by means of distributed and centralized fusion algorithms at the various levels of the network through local aggregation and fusion. Final goal is to achieve high confidence information regarding to the operation of the industrial system to be monitored aimed at decision support, optimization and planning. *Our specific application focus is on wireless sensor networks (WSN) and unmanned aerial vehicles (UAV) collaboration for joint monitoring of pipeline infrastructure scenarios.* The operational framework is that of multi-sensory intelligent environments which combine elements of computing, communication, control and cognition under an unifying novel paradigm. The particular nature of such large scale distributed transport systems poses a natural fit to the common characteristics of both WSNs and UAVs, namely: autonomous operation, communication, and local processing along with robust networking protocols. The main types of sensing and information processing architectures suitable for this type of application were previously defined as: conventional hierarchical, flat heterogeneous mobile sensor network structure, hybrid approaches [1]. Figure 1 presents a graphical illustration of such a collaborative system for pipeline surveillance.

A sensor fusion architecture, aimed specifically at identifying pipeline leaks and perimeter breaches is found appropriate, along with associated mechanisms aimed at fault tolerance and recovery, adapted to these architectures. The implementation can take multiple forms, combining the outcome of local sensor network event detection algorithms with image processing output. In summary, the following challenges are identified:

- Design of an integrated framework for information processing with application to large scale oil transport systems, based on collaborative monitoring through WSNs and UAVs;
- Evaluation and implementation of data aggregation and multi-sensor integration to reduce communication burden and uncertainty across the network by leveraging local computation resources;
- Combining scalar and multi-dimensional (image, video) data for high level management of situational awareness in pipeline infrastructures;
- Dynamic optimal tuning of hardware configuration and software/protocols adapted to event detection in large scale monitoring;
- Achieve simulation results as well steps toward actual deployment, to evaluate the feasibility of the proposed solution in terms of reliability and energy efficiency [2].

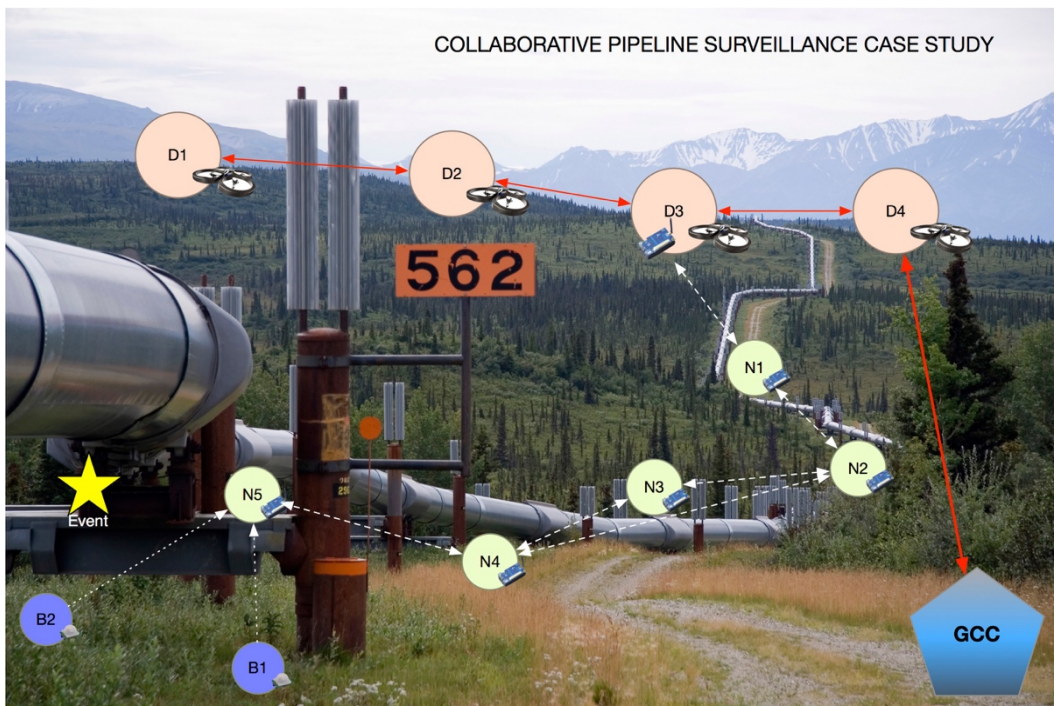

Fig. 1. Operation scenario and system architecture

The rest of the paper is structured as follows. Section two discusses recent related work relevant for real-world deployments of industrial wireless sensing solutions. Section three presents a detailed overview of the proposed system architecture for communication, data processing and interoperability. Discussion regarding three sensor fusion methods, applied to data processing from multiple sources is carried out in section four. Section five concludes the paper and opens up directions for future investigation.

## 2. Related work

Several works have recently approached the emergence of ubiquitous networked embedded devices in industrial context. A thorough study surveying one of the key issues hampering adoption of embedded computing and communication device in an industrial setting is performed in [3]. This is related to dependability at the component, node and system levels. WSN-specific attacks are classified and their impact on health monitoring and machine diagnostics is investigated. A framework for network and data management, I3WSN, is introduced and evaluated in [4]. The authors detail the system architecture, with multiple local WSNs collecting data at the field level which then aggregated in a global control center (GCC). Leveraging web technologies enables easy queries, service implementation and data visualization and storage. The prominent application to showcase the usefulness in an industrial setting is event detection for safety in risk management. The paper does not handle actual hardware implementation and constraints of real-world deployments and it is limited to an indoor environment and static system architecture with centralized data fusion at the top level.

An key issue in WSN deployments for manufacturing and industrial settings is localization for static and mobile entities. Conventionally this has been done by performing location inference by means of exploiting propagation behavior over the radio communication link. Basic approaches use the received signal strength indicator (RSSI) in relation to known fixed-position anchors to carry out trilateration for localization. More advanced methods include time of arrival (ToA), time difference of arrival (TDoA) or angle of arrival (AoA), depending on the case the need for more precise clock synchronization or more complex hardware might arise. In [5] mobile object localization is performed using ultrasound beacons in a manufacturing environment. Both active, which involves node to anchor distance ranging, and passive, anchor to node positioning, are covered. As trilateration may incur significant positioning error an intelligent method to improve the estimates is proposed, making use of an artificial neural network (ANN) to compensate signal noise and propagation effects. The scenario however is limited to a basic circular trajectory over a limited time-span and does not account for multiple node positioning. A more in-depth look at the radio channel modelling for low-power wireless communication is handled in [6]. The authors aim at deriving an optimal positioning strategy for facility sensor networks i.e. industrial wireless sensor networks for manufacturing plants, by statistical data analysis which can accommodate the multiple sources of interference, multi-path propagation effects and sensor faults and redundancy constraints. By using a fuzzy sensor validation and fusion (FUSVAF) method [7], confidence indicators are assigned to individual sensor readings which leads to an improved overall estimate of the desired parameter. Additionally, at the node level data can be pre-filtered in order to exploit on-board processing capabilities of the node. The optimal partitioning scheme can be used for various deployments and by assimilating UAVs as mobile nodes, they can be included in the global data processing framework running algorithms suited to the computing and communication resources available locally.

Of direct relevance to industrial pipeline monitoring applications, several systems have been proposed, implemented and evaluated. Initial experimental work was carried out for leakage detection and prevention for water pipelines. A system was proposed in [8] which uses the Intel Mote platform for pressure and acoustic/vibration measurement along the pipes to infer leakages. Signal processing is carried out locally and consists of frequency analysis for extraction and thresholding of Haar-wavelet coefficients. Upon surpassing the threshold an alert is conveyed towards the sink node. High energy content in certain frequency bands is also associated to possible leaks and correlated across neighboring nodes. A laboratory test-bed for system design is described in [9] which uses force-sensitive resistors (FSR) for non-invasive relative pressure monitoring. The sensor values are reported by the wireless sensor network and analyzed centrally to detect potential leaks. Oil-head pipeline monitoring with wireless embedded sensors is presented in [10]. As novel element, nodes are self-sufficient, relying of energy harvesting by means of thermoelectric generators for power. The gradient between hot steam pipes flowing into the well and external environment is exploited and proven to assure ample power reserves for nodes operating a conventional sense-store-transmit application.

### 3. System architecture

*The first level* includes the individual sensor nodes which measure pipeline pressure to indicate possible leaks, detect presence by means of PIR and magnetic sensors and also measure ambient parameters such as temperature, humidity, etc. These are small embedded computing and communication devices which communicate wirelessly through a low-power radio interface and form a mesh network to reliably convey data towards the sink. *The second level* is the sink/cluster-head level. Multiple small scale wireless sensor networks are integrated at this level to assemble the large scale monitoring system. The central point of the network is a node with more computing and communication resources which acts as network coordinator for the first level but also communicates with its peers (the coordinators of neighboring networks) over a longer range communication interface. The upper, *third level of the framework*, includes the interaction between the cluster heads and an unmanned aerial vehicle (UAV) tasked with a support role in the framework. Its role is two-fold, first it has the ability to relay WSN data to the central point of the monitoring system, called ground control center (GCC), but it is also used to collect visual information in the form of static and dynamic images to enhance data collection and event detection at the ground level. One relevant example is dispatching the UAV to validate a perimeter breach along the pipeline by target tracking and identification.

A main design option is to establish a decentralized data fusion architecture with the goal of efficient event detection and situation awareness. As background, such an approach offers some notable advantages concerning mainly communication and scalability. By leveraging the local computing resources of the nodes, the data processing load is distributed. This leads to reducing the impact of communication and signal processing bottlenecks in the overall system. Also, as large scale monitoring systems can be composed of hundreds to thousands of nodes, scalability is an important concern. By implementing effective processing distribution mechanisms, new entities (nodes, networks or UAVs) can be easily added to the system without significant additional overhead.

From a sensor integration perspective, by adhering to the three layer model previously introduced, several data processing operations are assigned to each level. On the individual node level, some basic filtering e.g. low-pass filter for noise suppression and thresholding can be applied on the raw data before communication. It has been previously estimated that, from an energy efficiency point of view, the energy to transmit 1 bit of data over a low-power wireless link equates to between 1000 - 3000 microcontroller operations, so that there is a convincing incentive to process data on-board. The second, higher level, for processing takes place at the network coordinator/cluster head level. As this node is the central collection point for data generate by the rest of the nodes, it is able to perform aggregation and fusion across time and domains, to the extend that it's computing resources allow. This ranges from basic aggregation operators such as COUNT, AVG, MAX, MIN, etc. to more complex functions that implement data fusion methods. As the role of the central node is both down-stream, coordination of the local network and up-stream, among peers and towards the UAV system. One novel implementation is the Opal sensor node [11] which uses two radios for communication in the 2.4GHz and 900 MHz bands and exploits thus communication diversity. Finally, at the top level, scalar data collected and processed by the ground sensor networks can be used to enhanced static and video surveillance of the pipeline infrastructure by a dedicated UAV. The two sources are fused at the ground control center (GCC) where humans also enter the loop for decision making. Associated to the GCC, the central coordination and fusion node of the wireless sensor network (NCW), provides an interface between the GCC and the deployed ground sensors. It provides the required computing resources for fusion algorithm implementation and bi-directional communication through a low-power radio interface. A diagram of the system architecture is shown in Figure 2.

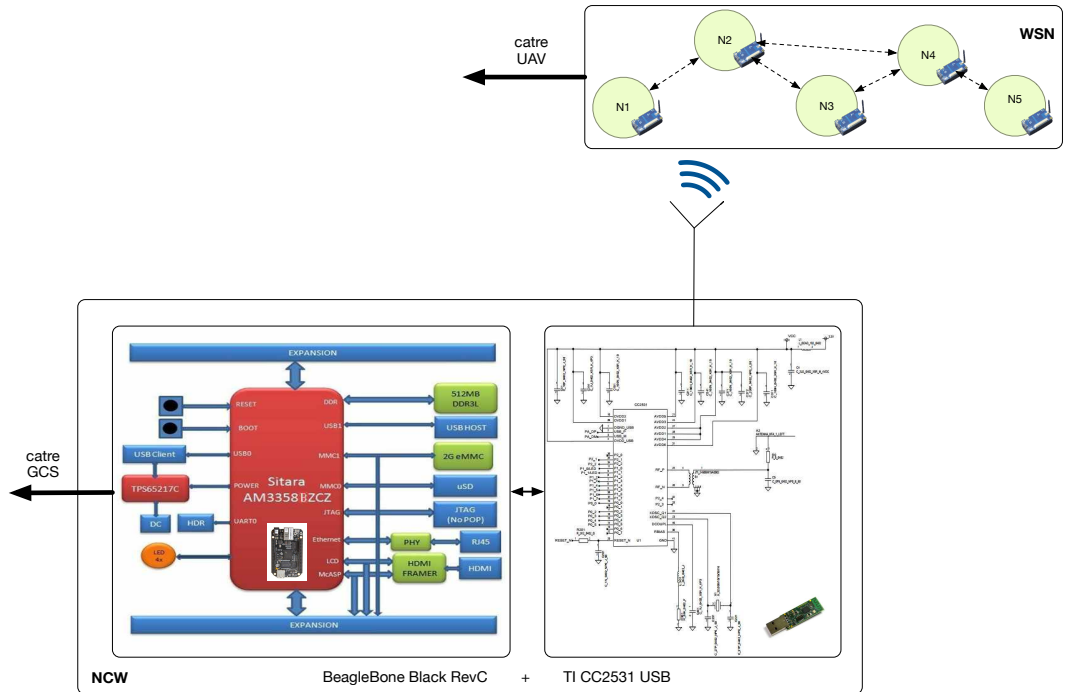

Fig. 2. Design of WSN fusion gateway

    System integration has been realized by combining off-the-shelf components, namely a BeagleBone Black embedded development board running Linux along with a TI CC2531 USB radio module, acting as radio base station for the IEEE 802.15.4 compliant ground sensor network. The Beaglebone leverages a Cortex-A8 processor at 1GHz and 512MB DDR3 memory and 4GB on-board flash along with a suite of connectivity options for interfacing, such as: USB, Ethernet, HDMI. By providing open-source software and hardware it is a very flexible solution. The radio module offers 2.4GHz ISM connectivity and mesh networking support under ContikiOS and supports IPv6 networking for the sensor network through 6LoWPAN protocols. The NCW is responsible for field data collection in raw or aggregated formats and performs sensor fusion according to the mission configuration. It interfaces through the serial USB port towards the GCC where the operator can leverage this data in an augmented reality environment for situational awareness.

## 4. Sensor fusion methods

Three sensor fusion methods were chosen for this analysis: Kalman filtering, Fuzzy Sensor Validation and Fusion (FUSVAF) and consensus-based. We briefly review their theoretical background along with simulation and experimental results in connection to our proposed application scenario.

- Extended Kalman filter (EKF)

The extended Kalman filter represents a generalization of the conventional Kalman filter (KF) towards non-linear systems. In addition to KF, it involves a linearization step at each cycle, which is based around the most recent state estimate. Is is important to note that, as opposed to KF, EKF is not an optimal filter and it lacks convergence where the approximation of the linear model is not good enough on the entire uncertainty interval. Taking the case of the non-linear discrete system:

$$x_{k+1} = f_k(x_k) + w_k \qquad (1)$$
$$y_k = h_k(x_k) + v_k \qquad (2)$$

with $v_k$ and $w_k$ zero mean, Gaussian noise and covariance matrices $R_k$ and $Q_k$ respectively. We denote $F(k)$ and $H(k)$ the Jacobian matrices of the functions $f$ and $h$.

$$F(k) = \nabla f_k|_{\hat{x}(k|k)} \qquad (3)$$
$$H(k+1) = \nabla h|_{\hat{x}(k+1|k)} \qquad (4)$$

The extended Kalman filter algorithm is implemented using two stages of prediction and measurement update, as follows:

Prediction:
$$\hat{x}(k+1|k) = f_k(\hat{x}(k|k)) \qquad (5)$$
$$P(k+1|k) = F(k)P(k|k)F^T(k) + Q(k)$$

Measurement update:
$$\hat{x}(k+1|k+1) = \hat{x}(k+1|k) + K(k+1)[y_{k+1} - h_{k+1}(\hat{x}(k+1|k))] \qquad (6)$$

$$K(k+1) = P(k+1|k)H^T(k+1)[H(k+1)P(k+1|k)H^T(k+1) + R(k+1)]^{-1} \qquad (7)$$
$$P(k+1|k+1) = [I - K(k+1)H(k+1)]P(k+1|k) \qquad (8)$$

Simulation results are presented in Figure 3. It illustrates the output of an uni-dimensional EKF for a stream of 20 raw measurements of a ground sensor. The initial estimates for process and measurement noise variance, $q$ and $r$, can be tuned to adjust system performance.

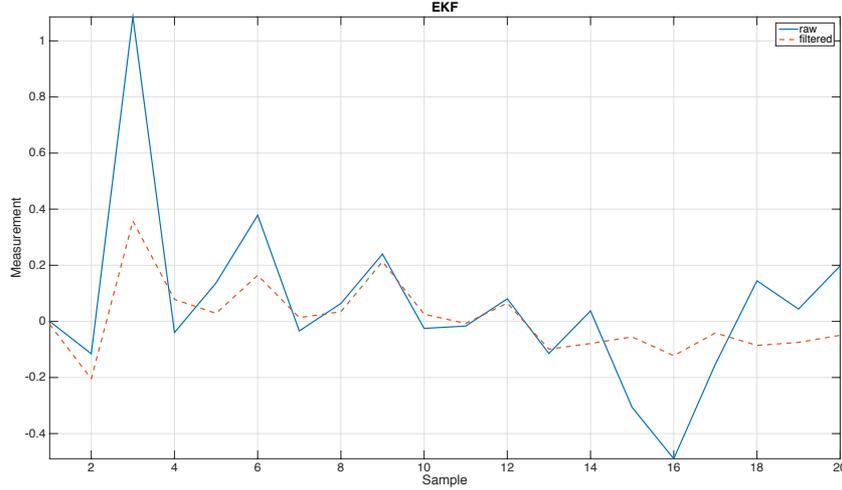

Fig. 3. EKF estimation of discrete measurements with r=0.1 and q=0.1

- Fuzzy Sensor Validation and Fusion (FUSVAF)

FUSVAF technique was implemented in [12] for sensor drift correction and previously applied in [13] for UAV altitude estimation based on barometric and GPS readings. It assumes constraining measurements from different sources through a dynamically adjusted validation gate. The validation gate can take the shape of a piece-wise bell curve, whose margins of are determined and updated by means of a fuzzy reasoning process. The confidence values are assigned by:

$$\sigma = \begin{cases} 0, & z < v_l \\ \dfrac{e^{-\left(\frac{\hat{x}-z}{a_l}\right)^2} - e^{-\left(\frac{\hat{x}-v_l}{a_l}\right)^2}}{1 - e^{-\left(\frac{\hat{x}-v_l}{a_l}\right)^2}}, & v_l < z \leq \hat{x} \\ \dfrac{e^{-\left(\frac{\hat{x}-z}{a_r}\right)^2} - e^{-\left(\frac{\hat{x}-v_r}{a_r}\right)^2}}{1 - e^{-\left(\frac{\hat{x}-v_r}{a_r}\right)^2}}, & \hat{x} < z \leq v_r \\ 0, & z > v_r \end{cases} \quad (9)$$

The estimated fused value, $\hat{x}_f$, is computed using the equation:

$$\hat{x}_f = \frac{\sum_{i=1}^{n} z_i \sigma(z_i) + \frac{\alpha \hat{x}}{\omega}}{\sum_{i=1}^{n} \sigma(z_i) + \frac{\alpha}{\omega}} \quad (10)$$

where $z_i$ are the measurements, $\sigma$ are the confidence values, $\alpha$ is an adaptive parameter which represents system state, $\omega$ is a constant scaling factor and $\hat{x}$ is the predicted value.

Figures 4 and 5 presents the outcome of the fusion method for experimental temperature and humidity data, collected from two ground sensor nodes.

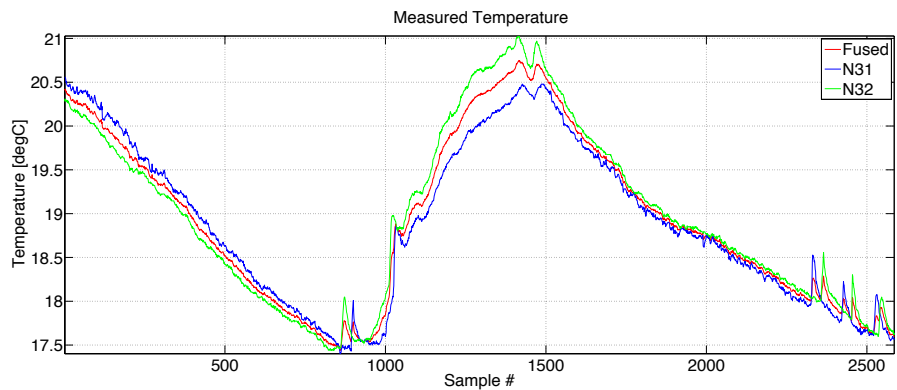

Fig. 4. Fused temperature measurements

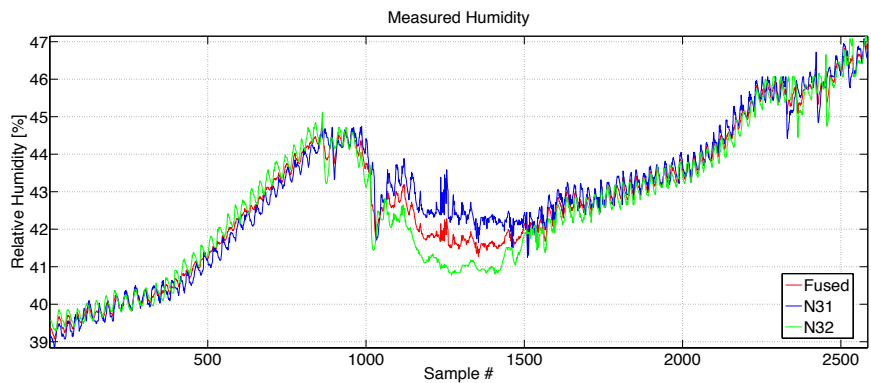

Fig. 5. Fused humidity measurements

- Consensus-based methods

Consensus-based methods assume a distributed optimization process across a local cluster of sensing entities. The theoretical background and proofs have been extensively discussed in [14,15]. Figure 6 represents the outcome of a consensus-based mechanism implementation where the mean squared error (MSE) between individual estimates is considered as metric and stop-condition for the iterative algorithm.

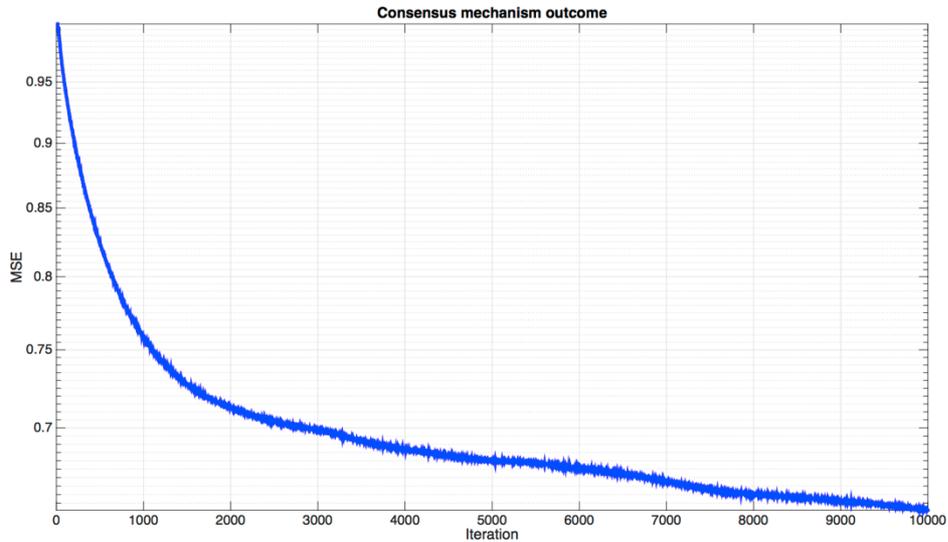

Fig. 6. MSE iterative consensus

### 5. Conclusions

The paper discussed the application of multi-level sensor integration to large scale monitoring systems based on collaborative WSN and UAV entities. The chosen application, pipeline infrastructure systems offer signification potential for real-world implementations due to it's key characteristics, scale and critical infrastructure designation. By studying the three methods for sensor fusion methods at the local and network levels, one of the main findings is that they can be used in a complementary fashion to achieve the best outcome. The concurrent use of the three algorithms can be implemented as follows:

- Kalman filtering for on-node pre-processing of raw data streams;
- FUSVAF at the cluster-head level for fault detection and weighted fusion of pre-processed data;
- Consensus-based methods, for on-demand, in-network pervasive information where the decision is distributed across the heterogeneous computing and communication entities.

Current and future work is aimed at deploying an integrated functional model of the monitoring systems, including WSN and UAV components and proving it's applicability in a real scenario.

### Acknowledgement

The work has been partially funded by the Sectoral Operational Programme Human Resources Development 2007-2013 of the Ministry of European Funds through the Financial Agreement POSDRU/159/1.5/S/134398.